\documentclass[manuscript,nonacm]{acmart}
\setcopyright{none}
\usepackage{listings}
\usepackage{tabularx}
\usepackage{placeins}
\lstdefinestyle{projectdocument}{
  basicstyle=\ttfamily\footnotesize,
  columns=fullflexible,
  keepspaces=true,
  breaklines=true,
  breakatwhitespace=true,
  breakindent=1em,
  showstringspaces=false,
  frame=single,
  framerule=0.3pt,
  framesep=5pt,
  rulecolor=\color{black!35},
  xleftmargin=6pt,
  xrightmargin=6pt,
  aboveskip=\medskipamount,
  belowskip=\medskipamount,
  captionpos=b,
  tabsize=2,
  literate={→}{{$\rightarrow$}}1 {—}{{---}}3 {–}{{--}}2
           {§}{{\S}}1 {≤}{{$\leq$}}1 {…}{{\ldots}}3
}
\lstdefinestyle{portcomparison}{
  basicstyle=\ttfamily\footnotesize,
  keywordstyle=\bfseries,
  commentstyle=\color{black!65},
  columns=fullflexible,
  keepspaces=true,
  breaklines=true,
  breakatwhitespace=false,
  showstringspaces=false,
  frame=single,
  framerule=0.3pt,
  framesep=4pt,
  rulecolor=\color{black!35},
  xleftmargin=5pt,
  xrightmargin=5pt,
  aboveskip=\smallskipamount,
  belowskip=0pt,
  tabsize=2
}
\makeatletter
\newenvironment{listing}[1][t]
  {\@float{lstlisting}[#1]}
  {\end@float}
\makeatother
\AtBeginDocument{%
  }

\setcopyright{acmlicensed}
\copyrightyear{2026}
\acmYear{2026}
\acmDOI{XXXXXXX.XXXXXXX}
\acmConference[Conference acronym 'XX]{Make sure to enter the correct
  conference title from your rights confirmation email}{June 03--05,
  2027}{Woodstock, NY}

\begin{document}
\raggedbottom
\title{Migrating a Legacy Scientific Application with an AI Coding Agent: NICAM from Fortran to Kokkos}

\author{Xuanzhengbo Ren}
\email{xuanzhengbo.ren@cc.u-tokyo.ac.jp}
\orcid{1234-5678-9012}
\correspondingauthor
\affiliation{%
  \institution{Information Technology Center, The University of Tokyo}
  \city{Kashiwanoha}
  \state{Chiba}
  \country{Japan}
}

\author{Takashi Shimokawabe}
\affiliation{%
  \institution{Information Technology Center, The University of Tokyo}
  \city{Kashiwanoha}
  \state{Chiba}
  \country{Japan}
}

\author{Hisashi Yashiro}
\affiliation{%
  \institution{National Institute for Environmental Studies}
  \city{Onogawa, City of Tsukuba}
  \state{Ibarak}
  \country{Japan}
}

\renewcommand{\shortauthors}{Ren et al.}

\begin{abstract}
Migrating from Fortran to C++/Kokkos broadens access to hardware accelerators through backends such as CUDA and HIP while offering greater flexibility in software design. 
However, legacy Fortran applications often have large codebases, making manual porting time-consuming and costly. 
Recent advances in large language models (LLMs) and AI coding agents have made large-scale code migration increasingly feasible. 
In this study, we ported NICAM (Nonhydrostatic ICosahedral Atmospheric Model), a legacy Fortran climate model accelerated with OpenACC, to C++/Kokkos with the assistance of Claude Code. 
The agent translated the code directly from Fortran to C++/Kokkos under human review.
Approximately 140,000 lines of Fortran were analyzed and ported into 35,000 lines of C++/Kokkos, covering the dynamical core and seven physics packages. 
A 100-member ensemble comparison with the original Fortran model showed that the port is statistically consistent with the original.
Performance was evaluated on Miyabi-G, an NVIDIA GH200 cluster, and Sirius, an AMD MI300A cluster. 
Compared with the Fortran/OpenACC baseline on GH200, the port achieved application speedups of $1.71\times$ on GH200 and $1.54\times$ on MI300A.
The CUDA build achieved a GPU kernel-time speedup of $1.38\times$ on GH200.
After normalization by peak memory bandwidth, the HIP build on MI300A achieved $0.96\times$ the kernel performance of the Fortran/OpenACC baseline.
\end{abstract}





\maketitle

\section{Introduction}
\label{sec:introduction}
High-performance computing (HPC) systems span a diverse range of processor and accelerator architectures, with different requirements for parallel execution and memory access.
Scientific applications must adapt to this diversity while remaining maintainable over development lifetimes that often outlast individual hardware generations.
Performance portability is therefore an important goal.
A shared implementation should run efficiently across target architectures without requiring a separate application codebase for each platform.
Kokkos addresses this goal through C++ abstractions for parallel execution, memory spaces, and multidimensional data layouts~\cite{edwards2014kokkos}.
Its CUDA and HIP backends provide a route to NVIDIA and AMD GPUs from a common source codebase~\cite{trott2022kokkos3}.

Adopting such a framework is a substantial undertaking for mature scientific applications because many have accumulated large Fortran codebases over decades.
These codebases combine numerical algorithms, physical parameterizations, communication routines, and input/output infrastructure.
Migrating these applications to C++/Kokkos requires both language translation and careful mapping of data structures and parallel execution while preserving numerical behavior.
The HOMMEXX atmospheric dynamical core illustrates the feasibility of this approach and the verification and performance engineering involved~\cite{bertagna2019hommexx}.
The expertise and time required for manual migration remain a barrier to porting a complete application.

Recent advances in large language models (LLMs) and coding agents offer a way to assist this work.
With access to development tools, these agents can combine source-code translation with compilation, testing, and iterative correction.
Fortran-to-Kokkos studies have demonstrated this approach on scientific kernels~\cite{gupta2025agentic}.
Application-scale studies have also reported agent-assisted migration of FESOM2 and ORB5 to C++/Kokkos under expert supervision~\cite{koldunov2026fesom,sadr2026orb5x}.
Together, these studies demonstrate the potential of coding agents for scientific software migration.
For a production atmospheric model, the resulting implementation must preserve the behavior of coupled dynamical and physical processes while executing efficiently across target architectures.
We therefore investigate whether human-directed, agent-assisted migration can meet these requirements in an application-scale C++/Kokkos port.

We address this question using the Nonhydrostatic ICosahedral Atmospheric Model (NICAM), whose original dynamical framework was developed by Tomita and Satoh~\cite{tomita2004nicam}.
NICAM uses an icosahedral grid and nonhydrostatic equations to support global atmospheric simulations.
At sufficiently fine resolution, it explicitly represents deep convective circulations~\cite{satoh2014nicam}.
Its development over more than two decades has included substantial adaptation to successive supercomputers.
Yashiro et al.\ applied OpenACC to its dynamical core on TSUBAME2.5~\cite{yashiro2016performance}, and subsequent work demonstrated large-scale NICAM-based ensemble data assimilation on Fugaku~\cite{yashiro2020ensemble}.
Building on these efforts, we aim to make a production configuration of NICAM available on both NVIDIA and AMD GPUs through C++/Kokkos.
The C++ implementation also provides a basis for future tuning through Kokkos and, where needed, native CUDA or HIP interfaces.

In this study, we use Claude Code~\cite{anthropicClaudeCode}, primarily with the Fable~5 model, to translate NICAM directly from Fortran to C++/Kokkos under human direction.
The human defines the scope and constraints, reviews plans and structural decisions, and assesses progress at defined checkpoints.
The migration proceeds module by module, using the Fortran implementation as the numerical reference and retaining its organization to support comparison and maintenance.
Our scope covers the dynamical core and seven physics packages used in the selected production configuration.
Component-level checks support development, while ensemble consistency testing evaluates the completed port outside the agent loop.
This paper makes three contributions.
\begin{itemize}
  \item An application-scale port of NICAM's selected production configuration, translating approximately 140,000 lines of Fortran into about 35,000 lines of C++/Kokkos model source and covering the dynamical core and seven physics packages.
  \item A layered validation approach combining component-level reference comparisons and an ensemble-inspired consistency test for coupled atmospheric simulations.
  \item An experimental evaluation of performance on Miyabi-G (NVIDIA GH200) and Sirius (AMD MI300A), together with comparisons of 100-member ensembles from each implementation to assess agreement in simulated means and variability.
\end{itemize}

The remainder of this paper is organized as follows.
Section~\ref{sec:related-work} reviews related work on performance-portable scientific applications, agent-assisted code migration, and ensemble consistency testing.
Section~\ref{sec:workflow} presents our workflow, porting strategies, and verification protocol.
Section~\ref{sec:evaluation} reports correctness, performance, and cost results and discusses their limitations.
Section~\ref{sec:conclusion} concludes the paper and outlines future work.

\section{Related Work}
\label{sec:related-work}
Modernizing weather and climate models for heterogeneous hardware predates LLM-based code generation.
Kokkos abstracts parallel execution and data layout across architectures~\cite{edwards2014kokkos,trott2022kokkos3}, enabling applications such as LAMMPS to target multicore CPUs and GPUs~\cite{moore2019lammps}.
SCREAM uses C++/Kokkos for a global atmosphere model designed for cloud-resolving simulations~\cite{donahue2024scream}.
Its dynamical core, HOMMEXX, is a C++/Kokkos port of a Fortran atmospheric dynamical core that preserves the underlying numerical formulation~\cite{bertagna2019hommexx}.
These efforts demonstrate both the feasibility of application-scale performance portability and the engineering effort it requires.

Compiler directives and domain-specific languages offer additional routes.
Earlier NICAM work accelerated its Fortran dynamical core with OpenACC~\cite{yashiro2016performance}.
CLAW generates OpenACC or OpenMP code for Fortran physical parameterizations from a single-column abstraction~\cite{clement2019claw}.
Pace expresses the FV3 dynamical core and GFDL cloud microphysics in Python and uses GT4Py for architecture-specific code generation~\cite{dahm2023pace}.
These approaches delegate hardware adaptation to compilers and code generators, whereas agent-assisted migration helps construct the translated application, with Kokkos providing portability.

Early studies of LLMs for HPC focused on the quality of generated kernels and the effects of prompting.
Godoy et al.\ evaluated OpenAI Codex on numerical kernels across several parallel programming models, including Kokkos, and found that the generated code depended on both the programming model and the prompt~\cite{godoy2023codex}.
Nichols et al.\ introduced ParEval to evaluate generated parallel code across scientific computing tasks, using metrics that consider correctness and execution performance~\cite{nichols2024parallel}.
Such studies provide controlled settings for examining model capabilities, but successful kernel generation alone does not establish the correctness of a coupled scientific application.

Subsequent work incorporated compiler and execution feedback into iterative translation.
Fortran2CPP uses a dual-agent process to generate translation dialogues covering compilation, execution, unit testing, and error correction.
It then uses these dialogues to train translation models~\cite{chen2024fortran2cpp}.
Gupta et al.\ present a multi-agent workflow for translating Fortran kernels to Kokkos, with separate roles for translation, compilation, testing, debugging, and optimization~\cite{gupta2025agentic}.
These studies demonstrate how automated feedback can be organized during migration, with evaluations focused mainly on individual programs or kernels.

Recent studies have extended agent-assisted translation to larger scientific applications.
Koldunov et al.\ report a port of the FESOM2 ocean model under expert supervision, first translating the Fortran core to C and then moving to C++/Kokkos~\cite{koldunov2026fesom}.
Their validation includes comparisons with an intermediate C reference at the kernel and whole-model levels, together with comparisons against the original Fortran model in simulations lasting more than one model year.
The intermediate implementation separates language translation from the introduction of Kokkos parallelism.
Our NICAM port follows a direct Fortran-to-C++/Kokkos route and covers the dynamical core and seven physics packages of a production atmospheric configuration.

Sadr et al.\ present ORB5X, a direct C++/Kokkos translation of the Fortran ORB5 gyrokinetic particle-in-cell code~\cite{sadr2026orb5x}.
Their workflow combines incremental translation, preservation of source structure, developer review, and numerical checks.
Verification includes unit tests and comparisons of diagnostic fields from complete plasma simulations, and performance is evaluated on NVIDIA and AMD systems.
Our independently developed NICAM workflow shares these general practices.
The distinction in this study lies in the atmospheric application and its validation through component-level reference comparisons and ensemble statistics.

Ensemble testing has an established role in assessing changes to weather and climate software when bitwise agreement is unavailable.
Baker et al.\ introduced the CESM ensemble consistency test, which evaluates new simulations against the variability represented by an accepted reference ensemble~\cite{baker2015cect}.
Milroy et al.\ developed an ultra-fast variant using only nine model time steps~\cite{milroy2018ufcect}.
Zeman and Sch\"ar describe a methodology that compares ensembles from different model executables to detect changes in their statistical behavior~\cite{zeman2022ensemble}.
These studies motivate assessing a translated atmospheric model using the variability of perturbed simulations in addition to checking individual routines.
For NICAM, we combine component-level comparisons against Fortran reference outputs with an ensemble-inspired consistency test. 

\section{Coding Agent Workflow}
\label{sec:workflow}
\subsection{Overview}
\begin{figure}[t]
  \includegraphics[width=0.4\linewidth]{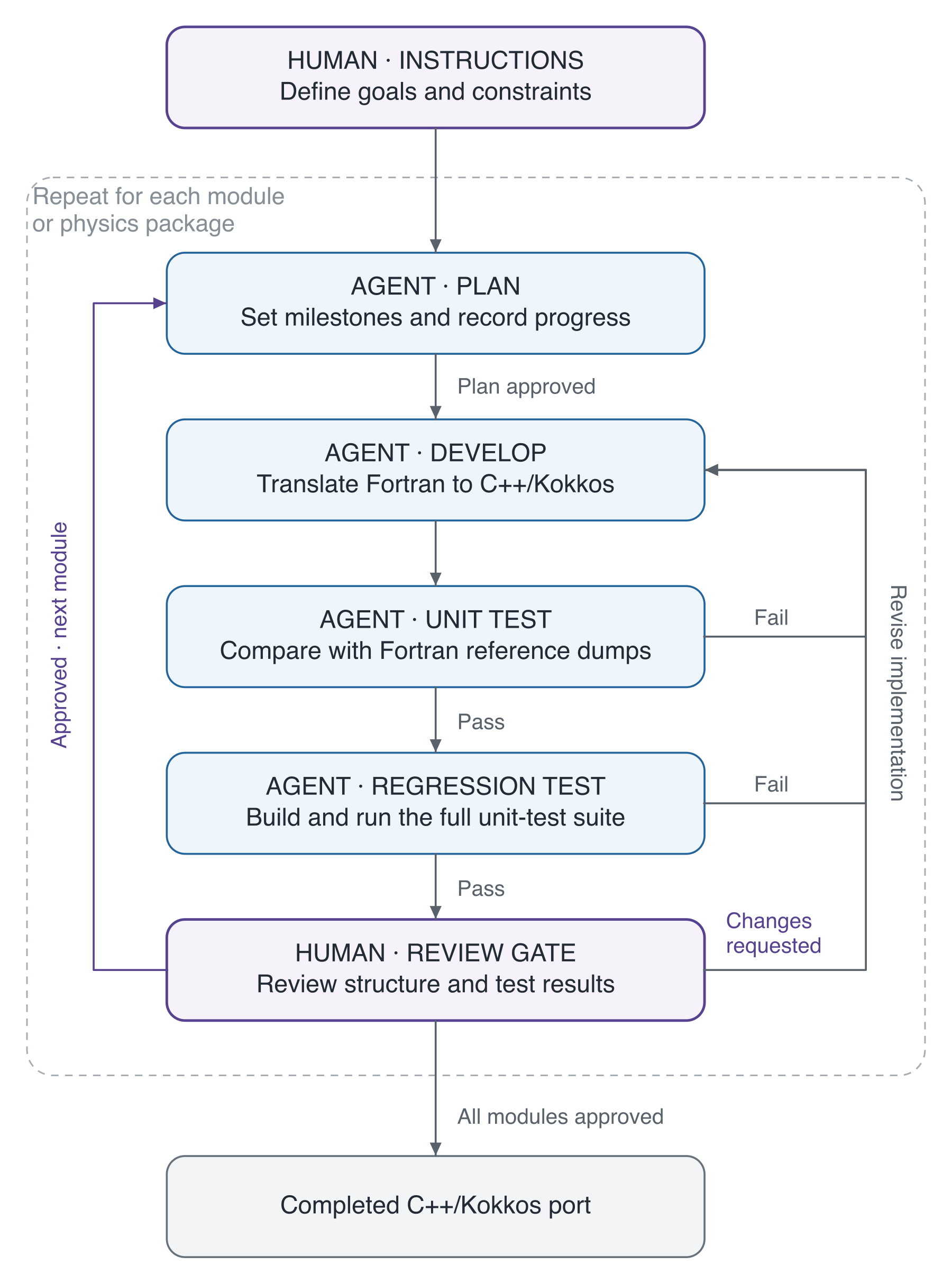}
  \caption{Coding agent workflow for porting NICAM to C++/Kokkos.}
  \label{fig:workflow}
  \Description{Coding agent workflow for porting NICAM to C++/Kokkos.}
\end{figure}

Figure~\ref{fig:workflow} summarizes the human-directed workflow used to port NICAM from Fortran to C++/Kokkos.
The human first defines the migration goals, requirements, and constraints.
In \texttt{plan} mode, the coding agent examines the Fortran source, clarifies the scope with the human, and proposes a porting plan with incremental milestones.
Development begins after the human approves the plan and proceeds one module or physics package at a time.
Within each milestone, the agent investigates the relevant routines and translates them directly into C++/Kokkos under the agreed constraints.

The agent then builds and tests the new implementation using component-level comparisons against the Fortran reference.
These tests execute the translated routines using inputs dumped from the Fortran model and compare their outputs with the corresponding reference dumps.
Once these checks pass, the agent rebuilds and runs the accumulated unit-test suite to detect regressions in previously implemented components.
A failure in either testing stage returns the workflow to development, where the agent revises the implementation and repeats the tests.

After both testing stages pass, the human reviews the test results and key structural decisions, including source organization and routine call order.
Short model runs of 30 or 60 time steps are also compared with the Fortran reference to check the integration of the new component.
If the human identifies a problem or requests further changes, the workflow returns to development.
Otherwise, the milestone is approved, progress is recorded, and the loop advances to the next module or physics package.
This process continues until all components in the selected configuration have been ported and approved.
Longer simulations and ensemble consistency testing subsequently assess the completed port outside the agent loop.

\subsection{Human Instruction and Planning}
We primarily used the Fable~5 model through Claude Code.
Human direction began with a concise kickoff document that specified the migration objective, immediate tasks, and operational constraints.
The document is reproduced in Appendix~\ref{sec:humaninst}.
The overall objective was to port NICAM from Fortran to C++/Kokkos, while the immediate task was to agree on a plan and develop a minimal executable version.
The instructions identified the local development environment, required the code to be built and run in Docker, and highlighted the original model's big-endian input/output.
They also required YAML configuration in place of Fortran namelists and set rules for file access, software installation, and GPU job execution.
The detailed breakdown of the port was left to the planning stage.

In Claude Code's \texttt{plan} mode, the agent examined the Fortran source and asked the human to resolve three planning choices.
First, we defined the initial executable as the dry dynamical core with physics disabled.
Second, we selected a standalone C++/Kokkos implementation to keep the target codebase self-contained and maintainable, rather than incrementally replacing routines within a mixed Fortran/C++ executable.
Third, we selected the single-header version of fkYAML for configuration parsing~\cite{fkyaml}.
These decisions established the initial scope, implementation strategy, and choice of configuration library before development began.

The agent recorded the agreed plan in \texttt{Porting\_Progress.md}.
The plan organized the dry dynamical core port into milestones that progressed from the build system and supporting modules to numerical operators, model state arrays, and the time-stepping driver.
A module map linked Fortran sources to proposed C++ files and reference checks, while milestone descriptions specified the planned work and completion criteria.
The document also included a preliminary roadmap for adding physics packages after the dry core and a log of decisions confirmed by the human.
Appendix~\ref{sec:portplan} presents selected excerpts from the initial plan generated on July~6, 2026.
The plan was subsequently refined as implementation and verification progressed.

\subsection{Mapping and Porting}
Fortran modules were mapped to C++ classes, with NICAM classes and functions organized within the \texttt{nicam} namespace.
The classes own model state arrays through Kokkos Views, and their instances are scoped so that the Views are destroyed before \texttt{Kokkos::finalize()}.
This makes data ownership and lifetime explicit.
Before launching a device kernel, member Views are copied into local handles that \texttt{KOKKOS\_LAMBDA} captures by value.
These shallow copies refer to the same allocations and allow the kernel to access the required arrays without capturing a pointer to the host-side owning object~\cite{kokkosLambdaDocs}.

The same organization supports isolated testing.
A unit test can construct only the objects required by a routine, populate them with dumped reference inputs, and execute the translated routine.
Separate object instances also allow multiple configurations to coexist without repeatedly resetting module-level global state.
Dependencies that were implicit in Fortran \texttt{use} statements become explicit in C++ interfaces.
For example, \texttt{Grd::setup} requires an \texttt{Adm} object, making the grid's dependency on the model topology visible to both the caller and the compiler.

Because individual NICAM modules can be large, the agent maintained working notes that mapped variables, routines, and code sections between the two implementations and recorded their callers and consumers.
These notes supported incremental translation and human review, although they did not constitute a formal dependency analysis.
The agent also tended to shorten variable names, which made some source correspondences less direct and required attention during structural review.

Listing~\ref{lst:flux-mapping} illustrates the mapping using an excerpt from the flux-convergence routine.
OpenACC compute regions become \texttt{Kokkos::parallel\_for} launches, while View definitions and explicit data transfers replace the corresponding array and data-management constructs.
The Kokkos panel combines the relevant type definitions, setup code, and kernel.
\texttt{LayoutLeft} retains the Fortran array ordering with the horizontal index \texttt{g} contiguous in memory.
The two-dimensional execution policy traverses \texttt{(k,g)}, with \texttt{g} varying fastest within a tile, and the outer loop processes regions indexed by \texttt{l}.
The temporary array \texttt{rhogw\_vmh(g,k,l)} is represented by the per-region scratch View \texttt{vmh(g,k)}.
Array indices are converted to zero-based indexing, and the Kokkos upper bounds are exclusive.

\begin{listing}[t]
  \begin{minipage}[t]{0.48\linewidth}
    \centering\small\textbf{(a) Fortran with OpenACC}\par
\begin{lstlisting}[style=portcomparison,language={[90]Fortran}]
subroutine src_flux_convergence( &
    rhogvx, rhogvy, rhogvz, rhogw, &
    grhog, ... )
  real(RP) :: rhogw_vmh( &
      ADM_gall,ADM_kall,ADM_lall)

  ! ... horizontal and vertical fluxes ...

  ! Total flux convergence
!$acc kernels pcopy(grhog) &
!$acc   pcopyin(rhogw_vmh,GRD_rdgz)
  do l = 1, lall
    do k = kmin, kmax
      do g = 1, gall
        grhog(g,k,l) = &
          - horizontal_flag*grhog(g,k,l) &
          - (rhogw_vmh(g,k+1,l) &
             - rhogw_vmh(g,k,l)) &
            * GRD_rdgz(k)
      enddo
    enddo
  enddo
!$acc end kernels

  ! ... pole-array handling ...
end subroutine src_flux_convergence
\end{lstlisting}
  \end{minipage}\hfill
  \begin{minipage}[t]{0.50\linewidth}
    \centering\small\textbf{(b) C++/Kokkos}\par
\begin{lstlisting}[style=portcomparison,language=C++]
// View and policy definitions
using Layout = Kokkos::LayoutLeft;
using ExecSpace =
    Kokkos::DefaultExecutionSpace;
using MemSpace = ExecSpace::memory_space;
template <class T = real_t>
using View3D =
    Kokkos::View<T***, Layout, MemSpace>;
using MDRange2 = Kokkos::MDRangePolicy<
    ExecSpace, Kokkos::Rank<2,
    Kokkos::Iterate::Right,
    Kokkos::Iterate::Right>>;
// MDR2(lo, hi) constructs MDRange2.

// Host-to-device copy at setup
rdgz_ = Kokkos::create_mirror_view_and_copy(
    ms, grd.rdgz);

// ... local View handles ...
for (int l = 0; l < adm.lall; ++l) {
  // ... horizontal and vertical fluxes ...

  // Total flux convergence
  Kokkos::parallel_for("src_fluxconv_total",
      MDR2({kmin, 0}, {kmax + 1, gall}),
      KOKKOS_LAMBDA(const int k,
                    const int g) {
        grhog(g,k,l) = -hflag*grhog(g,k,l)
            - (vmh(g,k+1) - vmh(g,k))
              * rdgz(k);
      });
}
\end{lstlisting}
  \end{minipage}
  \caption{Comparison of Fortran/OpenACC and C++/Kokkos implementations.}
  \label{lst:flux-mapping}
  \Description{Side-by-side source listings compare the Fortran OpenACC loop with Kokkos View definitions, initialization, and the equivalent parallel kernel.}
\end{listing}

Existing compiler directives also influenced the agent's translation choices.
OpenACC directives provided useful evidence about parallel regions and data movement.
However, the agent also attempted to introduce prefetching code in response to Fujitsu-specific \texttt{!\$OCL} directives in the source.
We reverted those changes and instructed the agent to avoid carrying such optimizations into the initial port.
This experience reinforced the need to review how source annotations were interpreted and to check proposed data transfers against the surrounding data flow.

\subsection{Verification Protocol}
Verification addressed three complementary questions through component-level reference tests, short integration runs, and an ensemble-inspired assessment of the completed model.
The first two tiers supported milestone acceptance within the workflow.
The third was conducted outside the agent loop to assess the statistical behavior of coupled simulations.

\textbf{Tier 1. Component-level reference tests.}
The tests used data from the Fortran simulations, including the atmospheric state and boundary conditions.
Before translating a routine, we dumped its inputs and outputs from the Fortran implementation into binary files.
The corresponding unit test supplied the same inputs to the C++/Kokkos routine and compared its outputs with the reference dumps.
The standard acceptance threshold was a relative error of at most $10^{-13}$ or an absolute error of at most $10^{-12}$, with many comparisons agreeing at approximately $10^{-15}$.
Exceptions for ill-conditioned intermediate quantities were documented, and their downstream consumers were checked under strict tolerances.
These tests exercised individual routine calls from a reference step, providing rapid feedback on translation errors without requiring a long model integration.
The accumulated test suite was rerun after each milestone to detect regressions.

\textbf{Tier 2. Human-reviewed short integration runs.}
At each milestone, we compared model fields and energy and mass budget diagnostics from runs of 30 or 60 time steps with the Fortran reference.
This stage checked initialization, coupling, and repeated execution for errors that isolated routine tests could miss.
For example, after the physics packages had passed their unit tests, an integrated run revealed a systematic discrepancy in surface downward longwave radiation.
The routine tests supplied the surface temperature directly from reference dumps, whereas the coupled model had to reconstruct it during surface initialization.
The corresponding reconstruction was missing from the port, so the unit tests did not expose the error.
The budget discrepancy prompted inspection and correction of the initialization path.
Human structural review complemented these numerical checks, particularly for dependencies and routine call order.

\textbf{Tier 3. Ensemble-inspired consistency testing.}
Atmospheric simulations amplify small numerical perturbations through chaotic dynamics, so agreement between individual trajectories becomes difficult to interpret as the integration proceeds.
The migration from Fortran/OpenACC to C++/Kokkos also changes the compiler and parallel programming model, which can alter the order of floating-point operations in reductions and make the resulting numerical differences harder to interpret.
We therefore assessed the completed port using an ensemble approach inspired by CESM-ECT~\cite{baker2015cect}.
Our test compares diagnostic quantities and their distributions directly using $z$-scores, without implementing the principal-component-based acceptance procedure of CESM-ECT.

Using the Fortran ensemble as the reference, the score for quantity $q$ and Kokkos member $m$ is
\begin{equation}
  z_{m,q}^{\mathrm{K}} =
  \frac{x_{m,q}^{\mathrm{K}}-\bar{x}_{q}^{\mathrm{F}}}
       {s_{q}^{\mathrm{F}}},
  \label{eq:ensemble-member-z}
\end{equation}
where $x_{m,q}^{\mathrm{K}}$ is the quantity summarized for that member, and $\bar{x}_{q}^{\mathrm{F}}$ and $s_{q}^{\mathrm{F}}$ are the Fortran ensemble mean and sample standard deviation.
The score measures the departure in units of the reference ensemble spread.
We generated the ensemble members by adding independent uniform temperature perturbations in $[-10^{-10},10^{-10}]$\,K to the initial state and used the same perturbed states for both implementations.
Each member ran for 10 model days.
For each member, we summarized budget diagnostics and global field statistics over the second half of the simulation, starting at model day~5.
Comparing the resulting distributions shows whether the Kokkos members exhibit departures similar to the variability within the Fortran ensemble.
Agreement between these distributions provides evidence of consistency for the evaluated diagnostics over the 10-day simulations.

\subsection{Final Application Layout}
This effort covered porting the dynamical core and physics packages, reviewing and revising the overall codebase, and conducting the ensemble consistency test.
Approximately 140,000 lines of Fortran were analyzed and ported to 35,000 lines of C++/Kokkos model source.

Several implementation differences account for the shorter C++/Kokkos source code.
The Fortran routines contain separate type and dimension declarations and multiple \texttt{do} and \texttt{end do} lines for nested loops, while Kokkos expresses multidimensional iteration more compactly through \texttt{MDRange}.
The port also replaces OpenACC and OpenMP directives with Kokkos constructs and uses C++ templates to avoid duplicating double-precision (DP) and single-precision (SP) implementations.
Branches outside the selected configuration were not ported.

The completed application combines NICAM's nonhydrostatic dynamical core with seven physics packages in the selected production configuration.
These comprise NSW6 cloud microphysics, MSTRNX radiation, Louis surface fluxes, MYNN2 turbulence, McFarlane orographic gravity-wave drag, MATSIRO land surface processes, and a mixed-layer ocean with sea ice.
The C++ source retains the broad organization of the Fortran model, separating common components, model state arrays, dynamics, physics, and the driver.
This organization helps developers locate corresponding components in the two implementations during maintenance and review.

The driver coordinates initialization, time integration, and calls to the selected physics packages.
Package selection retains the original \texttt{*\_TYPE} values, while the translated packages operate on model fields stored in Kokkos Views.
The prognostic state remains device-resident across time steps, and Kokkos provides the execution and memory abstractions used by the CUDA and HIP builds.
The port supports the dry dynamical core and the selected coupled configuration, with unported option branches rejected explicitly.

\begin{listing}[t]
  \begin{minipage}[t]{0.49\linewidth}
    \centering\small\textbf{(a) Fortran namelist}\par
\begin{lstlisting}[style=portcomparison,language={[90]Fortran}]
&ADMPARAM
  glevel = 7,
  rlevel = 0,
  vlayer = 78,
  ! ...
/

&TIMEPARAM
  INTEG_TYPE = 'RK2',
  LSTEP_MAX = 120,
  DTL = 30.D0,
  ! ...
/

&RUNCONFPARAM
  MP_TYPE = 'NSW6',
  opt_2moment_water = .false.,
  RD_TYPE = 'MSTRNX',
  SF_TYPE = 'LOUIS',
  TB_TYPE = 'MYNN2',
  GWD_TYPE = 'MCFARLANE',
  LAND_TYPE = 'MATSIRO',
  OCEAN_TYPE = 'MIXEDLAYER',
  ! ...
/
\end{lstlisting}
  \end{minipage}\hfill
  \begin{minipage}[t]{0.49\linewidth}
    \centering\small\textbf{(b) YAML}\par
\begin{lstlisting}[style=portcomparison,language={},morecomment={[l]{\#}}]
adm:
  glevel: 7
  rlevel: 0
  vlayer: 78
  # ...


time:
  integ_type: "RK2"
  lstep_max: 120
  dtl: 30.0
  # ...


runconf:
  mp_type: "NSW6"
  opt_2moment_water: false
  rd_type: "MSTRNX"
  sf_type: "LOUIS"
  tb_type: "MYNN2"
  gwd_type: "MCFARLANE"
  land_type: "MATSIRO"
  ocean_type: "MIXEDLAYER"
  # ...
\end{lstlisting}
  \end{minipage}
  \caption{Illustration of the Fortran namelist and C++/Kokkos YAML configuration.}
  \label{lst:configuration-mapping}
\end{listing}

Runtime configuration is read from YAML using fkYAML~\cite{fkyaml}.
Listing~\ref{lst:configuration-mapping} compares selected settings from the Fortran namelist and the corresponding YAML file.
Namelist groups become YAML mappings, such as \texttt{ADMPARAM} to \texttt{adm} and \texttt{RUNCONFPARAM} to \texttt{runconf}, with parameter names expressed in lowercase.
The settings retain their meanings for grid resolution, integration, and physics package selection.
Fortran logical and floating-point literals are expressed using YAML syntax, as illustrated by \texttt{.false.} becoming \texttt{false} and \texttt{30.D0} becoming \texttt{30.0}.

\section{Evaluation}
\label{sec:evaluation}
\subsection{Experimental Environment}
We evaluated the port on two supercomputers with different accelerator architectures.
Miyabi-G uses NVIDIA GH200 Grace Hopper Superchips and is operated by the University of Tokyo and the University of Tsukuba~\cite{miyabiSystem}.
Sirius uses AMD Instinct MI300A APUs and is operated by the Center for Computational Sciences at the University of Tsukuba~\cite{siriusSystem}.
Table~\ref{tab:platforms} summarizes the hardware, software, and principal build options.
On both systems, the port uses Kokkos~5.1.1 and C++20 in Release builds, with the CUDA backend on Miyabi-G and the HIP backend on Sirius.
The Fortran/OpenACC reference was evaluated on Miyabi-G.

\begin{table}[t]
  \caption{Experimental platforms and software configurations on Miyabi-G and Sirius.}
  \label{tab:platforms}
  \small
  \renewcommand{\arraystretch}{1.12}
  \begin{tabularx}{\linewidth}{@{}>{\raggedright\arraybackslash}p{0.21\linewidth}
      >{\raggedright\arraybackslash}X
      >{\raggedright\arraybackslash}X@{}}
    \toprule
    & \textbf{Miyabi-G} & \textbf{Sirius} \\
    \midrule
    Accelerators per node & $1\times$ NVIDIA GH200 & $4\times$ AMD MI300A \\
    CPU per accelerator & Grace, 72 Arm Neoverse V2 cores & 24 Zen~4 cores \\
    GPU architecture & NVIDIA H100 (Hopper) & AMD CDNA~3 \\
    HBM3 capacity & 96\,GB & 128\,GB, shared with the CPU \\
    Peak memory bandwidth & 4.02\,TB/s & 5.3\,TB/s \\
    Peak FP64 (vector) & 33.5\,TFLOP/s & 61.3\,TFLOP/s \\
    CPU--GPU connection & NVLink-C2C & Infinity Fabric \\
    Network per node & InfiniBand NDR, 200\,Gb/s & InfiniBand NDR, $4\times400$\,Gb/s \\
    \midrule
    Kokkos version & 5.1.1 & 5.1.1 \\
    C++ compiler & nvc++ 25.9 (NVHPC) & hipcc (ROCm 7.2.0, Clang 22) \\
    Fortran compiler & nvfortran 25.9 (NVHPC) & Not evaluated \\
    MPI & NVIDIA HPC-X (NVHPC 25.9)\newline CUDA-aware
        & Open MPI 5.0.10\newline ROCm-aware \\
    \addlinespace
    Kokkos build options
      & \path|Kokkos_ENABLE_CUDA=ON|\newline
        \path|Kokkos_ENABLE_OPENMP=ON|\newline
        \path|Kokkos_ENABLE_SERIAL=ON|\newline
        \path|Kokkos_ARCH_HOPPER90=ON|\newline
        \path|Kokkos_ARCH_ARMV9_GRACE=ON|\newline
        \path|Kokkos_ENABLE_TUNING=ON|
      & \path|Kokkos_ENABLE_HIP=ON|\newline
        \path|Kokkos_ENABLE_OPENMP=ON|\newline
        \path|Kokkos_ENABLE_SERIAL=ON|\newline
        \path|Kokkos_ARCH_AMD_GFX942=ON|\newline
        \path|Kokkos_ARCH_ZEN4=ON|\newline
        \path|Kokkos_ENABLE_TUNING=ON|\newline
        \path|Kokkos_ENABLE_HIP_RELOCATABLE_DEVICE_CODE=OFF| \\
    \addlinespace
    Fortran compiler flags
      & \texttt{-fast -Mfprelaxed -Mfpapprox -Kieee -Mvect=assoc -Mbyteswapio -Mflushz -mcmodel=large -acc}
        \path|-gpu=cc90,maxregcount:128|
      & --- \\
    \bottomrule
  \end{tabularx}
\end{table}

We used three grid configurations, all with 78 vertical layers.
The \texttt{gl07rl00Az78} configuration has approximately 56\,km horizontal spacing and 10 regions.
It was used during development, for the single-accelerator performance breakdown, and for the ensemble test.
For strong scaling, \texttt{gl07rl01Az78} retains the same horizontal resolution but partitions the domain into 40 regions.
The finer \texttt{gl09rl03Az78} configuration has approximately 14\,km spacing and 640 regions.
All experiments used double precision and the seven physics packages described above, with a 30\,s time step and radiation updates every 30 model minutes.

For the kernel-time breakdown, the radiation update cost was amortized over its 60-step interval.
NICAM's internal \texttt{rap} timers measured application elapsed time over all 120 steps, excluding I/O.
The kernel-time breakdown reports accumulated GPU kernel execution time.
For overall time per step and strong scaling, we used the model's \texttt{WCLOCK} timestamps over steps 20--119 in unprofiled runs, excluding initialization and final restart output.

\subsection{Correctness Results}
\begin{figure}[t]
  \centering
  \includegraphics[width=0.8\linewidth]{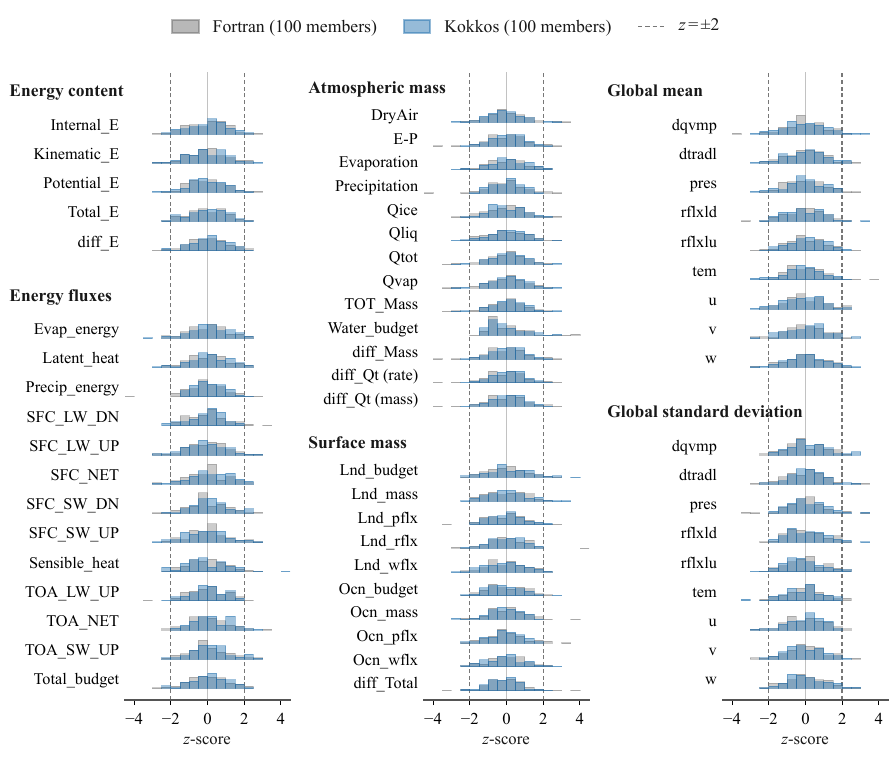}
  \caption{Ensemble consistency results for 59 quantities from two 100-member ensembles, summarized over model days~5--10.}
  \label{fig:ensemble-consistency}
  \Description{Three columns of overlapping gray and blue histograms compare 59 energy, mass, and global field diagnostics. Both ensembles have similar distributions around zero, with dashed reference lines at minus and plus two standard deviations.}
\end{figure}

Figure~\ref{fig:ensemble-consistency} compares the distributions of member $z$-scores for the 59 evaluated quantities over model days~5--10.
Fortran members were scored against the remaining Fortran members using leave-one-out means and standard deviations.
This prevents a reference member from contributing to the statistics used to score itself.
Each row shows the Fortran and Kokkos histograms for one quantity, with 100 members in each ensemble.
The distributions overlap substantially across energy and mass budget diagnostics and the global means and standard deviations of model fields.
Of the 5,900 member--quantity scores in each ensemble, 95.3\% lie within $|z|\leq2$ in both implementations.
The median number of quantities outside this interval is two per member in each ensemble.
This distributional agreement supports consistency of the port with the Fortran reference for the evaluated quantities over the 10-day simulations.

\subsection{Performance}
\begin{figure}[t]
  \centering
  \includegraphics[width=0.6\linewidth]{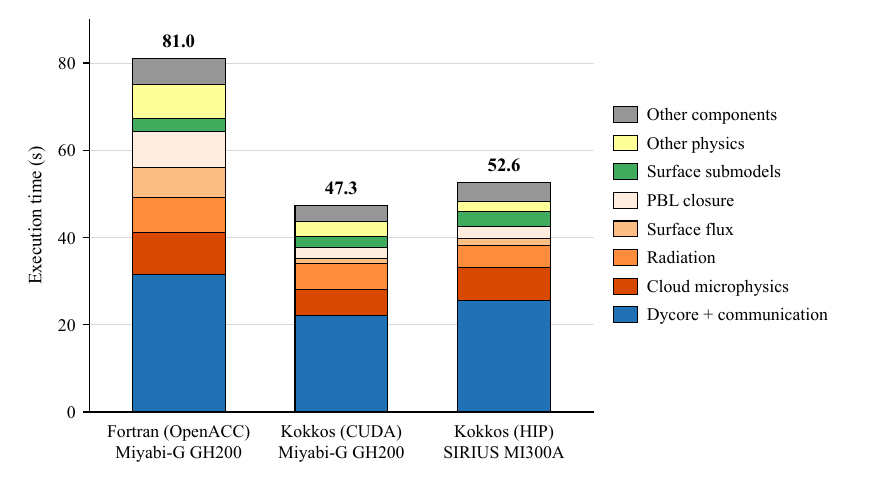}
  \caption{Application execution time over 120 time steps, measured by NICAM's \texttt{rap} timers.}
  \label{fig:rap-timer-breakdown}
  \Description{Three stacked bars compare the Fortran/OpenACC reference on Miyabi-G GH200, Kokkos/CUDA on Miyabi-G GH200, and Kokkos/HIP on Sirius MI300A. Total execution times are 81.0, 47.3, and 52.6 seconds, respectively. Eight segments distinguish dynamics and communication, physics packages, surface submodels, and other components.}
\end{figure}

Single-accelerator runs used one MPI rank per GPU or APU for 120 time steps on the \texttt{gl07rl00Az78} grid.
Figure~\ref{fig:rap-timer-breakdown} compares the Fortran/OpenACC reference on Miyabi-G with the CUDA and HIP builds of the Kokkos port on Miyabi-G and Sirius, respectively.
The reference takes 81\,s, compared with 47.3\,s for Kokkos on GH200 and 52.6\,s on MI300A.
These correspond to speedups of $1.71\times$ and $1.54\times$, both compared with the Fortran reference on GH200.
The port runs more slowly on MI300A than on GH200 despite MI300A's higher peak memory bandwidth and FP64 throughput.
The HIP build reuses the CUDA tile sizes in \texttt{MDRange}, which may limit performance on MI300A.
Differences between Kokkos's CUDA and HIP backend implementations may also contribute to the gap.

\begin{figure}[t]
  \centering
  \includegraphics[width=0.8\linewidth]{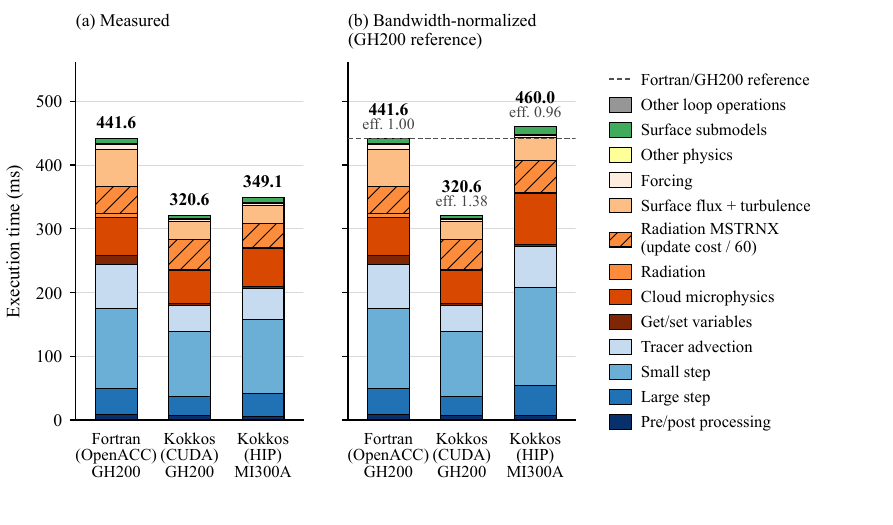}
  \caption{Device kernel execution time per model step, measured and normalized to GH200 peak memory bandwidth.}
  \label{fig:kernel-time-breakdown}
  \Description{Two panels compare stacked device kernel times for Fortran/OpenACC on GH200, Kokkos/CUDA on GH200, and Kokkos/HIP on MI300A. Panel (a) shows measured totals of 441.6, 320.6, and 349.1 milliseconds per step. Panel (b) normalizes the times to GH200 peak memory bandwidth, giving totals of 441.6, 320.6, and 460.0 milliseconds and relative efficiencies of 1.00, 1.38, and 0.96. A dashed line marks the Fortran/GH200 reference. Thirteen segments distinguish dynamical-core stages, physics, surface submodels, and other loop operations. The MSTRNX update cost is amortized over 60 steps and shown with hatching.}
\end{figure}

We also measured accumulated GPU kernel execution time using Nsight Systems with NVTX annotations on Miyabi-G and \texttt{rocprofv3} with \texttt{roctx} annotations on Sirius.
Kernel timings were collected over steps 20--119 to avoid startup effects.
Because the MSTRNX radiation update runs only once every 60 steps, its cost is divided by 60 before being added to the breakdown.
Figure~\ref{fig:kernel-time-breakdown}(a) reports the resulting kernel execution time per step.

The totals are 441.6\,ms for Fortran, 320.6\,ms for Kokkos on GH200, and 349.1\,ms for Kokkos on MI300A, giving kernel-time speedups of $1.38\times$ and $1.27\times$ compared with the Fortran reference.
The larger speedups in application elapsed time also reflect reductions in staging and synchronization overhead.
The Fortran physics implementation repeatedly gathers data from model arrays into per-region work arrays, with explicit \texttt{!\$acc wait} directives at these boundaries.
This per-region gather pattern may be inherited from the original CPU implementation.
The port instead operates on full-domain device arrays, reducing staging work and synchronization.
The kernel-duration sums include only time spent executing GPU kernels, while the \texttt{rap} measurements also capture the contribution of host waiting and launch overhead to application elapsed time.
The two breakdowns also cover different step ranges and treat the infrequent radiation update differently, so their totals cannot be directly subtracted to quantify overhead.

Figure~\ref{fig:kernel-time-breakdown}(a) shows that the dynamical core accounts for the largest absolute reduction in kernel time on GH200, with the total across the four blue segments decreasing from 243.3 to 178.9\,ms per step.
The Fortran implementation relies largely on compiler-generated kernels from \texttt{!\$acc kernels} regions, whereas the port specifies multidimensional iteration ranges and tile sizes explicitly.
For example, the Fortran offload in the vertically implicit solver is placed inside the region loop, while the port exposes work across regions through a multidimensional Kokkos execution policy.
This changes the amount of independent work available in each kernel launch.
The time spent in the ``Get/set variables'' kernels also decreases from 15.3 to 4.3\,ms on GH200.
The Fortran version copies model state arrays to and from per-region work arrays, performs prognostic-to-diagnostic conversions per region, and transfers the full pressure field to the host for an instability check.
The port reads the state array in place, converts the full field in one pass, and performs the check on the device.
The time spent in the ``Surface flux + turbulence'' kernels decreases from 58.8 to 28.5\,ms following the fusion of adjacent computations, which reduces launches and intermediate data traffic but makes the larger kernels harder to review and maintain.
However, improvements are not uniform across components.
For example, the amortized cost of the MSTRNX update increases from 41.9 to 46.3\,ms on GH200.

Figure~\ref{fig:kernel-time-breakdown}(b) compares the CUDA and HIP builds after normalization by peak memory bandwidth, using 4.02\,TB/s for GH200 and 5.3\,TB/s for MI300A.
Under this normalization, the HIP build on MI300A achieves approximately 70\% of the CUDA build's kernel performance on GH200.
Against the Fortran/OpenACC baseline on GH200, the corresponding performance ratios are $1.38\times$ for CUDA and $0.96\times$ for HIP.

\begin{figure}[t]
  \centering
  \includegraphics[width=0.7\linewidth]{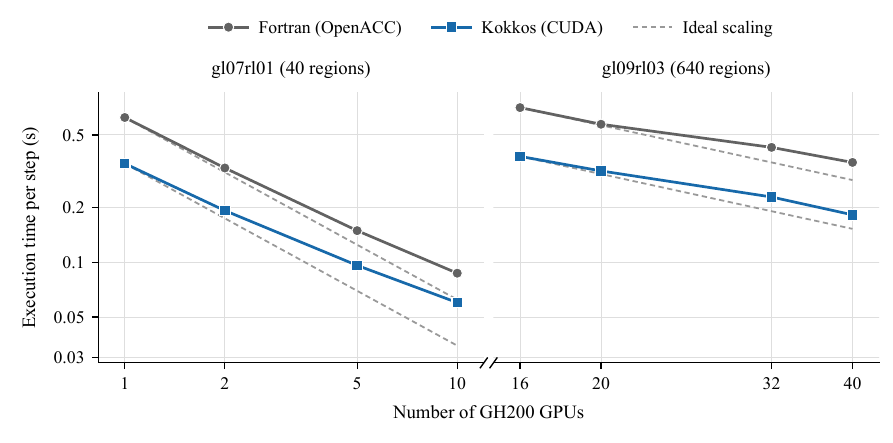}
  \caption{Strong-scaling execution time per step on Miyabi-G with four MPI ranks per GH200 GPU under MPS.}
  \label{fig:scaling-execution-time}
  \Description{Two equally wide adjoining horizontal ranges compare Fortran/OpenACC and Kokkos/CUDA execution times on a shared vertical scale. An axis break marks the different horizontal scales, and curves are not connected between the grids. For gl07rl01 with 40 regions, the GPU counts are 1, 2, 5, and 10, and the times decrease from 0.623 to 0.087 seconds for Fortran and from 0.348 to 0.060 seconds for Kokkos. For gl09rl03 with 640 regions, the GPU counts are 16, 20, 32, and 40, and the times decrease from 0.706 to 0.353 seconds for Fortran and from 0.381 to 0.182 seconds for Kokkos. All points use four MPI ranks per GPU under MPS.}
\end{figure}

To assess performance beyond a single accelerator, we conducted strong-scaling experiments on Miyabi-G with four MPI ranks per node sharing one GH200 GPU through MPS for both implementations.
Figure~\ref{fig:scaling-execution-time} compares execution time per step for two fixed problem sizes.
The \texttt{gl07rl01Az78} grid has 40 regions at approximately 56\,km spacing and uses 1, 2, 5, and 10 nodes.
The \texttt{gl09rl03Az78} grid has 640 regions at approximately 14\,km spacing and uses 16, 20, 32, and 40 nodes.
Each point represents one unprofiled run, with the time per step calculated from the 100-step \texttt{WCLOCK} window, excluding initialization and final restart output.

For \texttt{gl07rl01Az78}, increasing the GPU count from 1 to 10 reduces the Fortran execution time from 0.623 to 0.087\,s per step and the Kokkos time from 0.348 to 0.060\,s per step.
The port remains faster at every tested GPU count, although the gap narrows as the local workload decreases.
At 10 nodes, each of the 40 MPI ranks owns only one region.
Both implementations depart further from ideal inverse scaling as the GPU count increases, with a larger relative departure for the port.

For \texttt{gl09rl03Az78}, increasing the GPU count from 16 to 40 reduces the Fortran time from 0.706 to 0.353\,s per step and the Kokkos time from 0.381 to 0.182\,s per step.
This grid retains four regions per MPI rank at the largest configuration.
The two implementations follow similar scaling trends over the measured range, while the port maintains a lower execution time throughout.

\subsection{Token Cost and Human Effort}
We used Claude Code under a subscription plan and asked the agent to summarize token usage from the counters recorded in its session logs.
Table~\ref{tab:token-cost} reports the overall usage and its estimated equivalent cost at API list prices.
This includes tokens used not only for porting and verifying NICAM but also for generating scripts for plotting and data analysis.

Uncached input is processed at the full input rate and accounts for only a small fraction of usage because most conversation context is cached.
Cache writes store prompt content for later reuse, including new instructions, tool results, and previous replies.
For Fable~5, the ledger uses cache-write rates of $1.25\times$ the input rate for a five-minute lifetime and $2\times$ for a one-hour lifetime.
Cache reads reuse earlier context at $0.1\times$ the input rate.
Reusing the same context across calls accounts for the 3.88 billion cache-read tokens, even though the human did not enter the same material repeatedly.
Output tokens count newly generated text, tool calls, and reasoning and are charged at $5\times$ the input rate for Fable~5.
The overall API list-price estimate is approximately USD~5,589.

\begin{table}[t]
  \caption{Token usage and estimated API list-price cost for the port and analysis.}
  \label{tab:token-cost}
  \centering
  \small
  \renewcommand{\arraystretch}{1.1}
  \begin{tabular*}{0.7\linewidth}{@{\extracolsep{\fill}}lrrrr@{}}
    \toprule
    Counter & Tokens (M) & \shortstack[r]{Share of\\tokens (\%)} &
      \shortstack[r]{List price\\(USD)} & \shortstack[r]{Share of\\cost (\%)} \\
    \midrule
    Uncached input & 0.4 & $<0.1$ & 3.7 & 0.1 \\
    Cache write & 127.3 & 3.2 & 1,887.9 & 33.8 \\
    Cache read & 3,883.2 & 96.5 & 3,098.9 & 55.4 \\
    Output & 13.9 & 0.4 & 598.7 & 10.7 \\
    \midrule
    Total & 4,024.8 & 100.0 & 5,589.2 & 100.0 \\
    \bottomrule
  \end{tabular*}
\end{table}

\begin{figure}[t]
  \centering
  \includegraphics[width=0.7\linewidth]{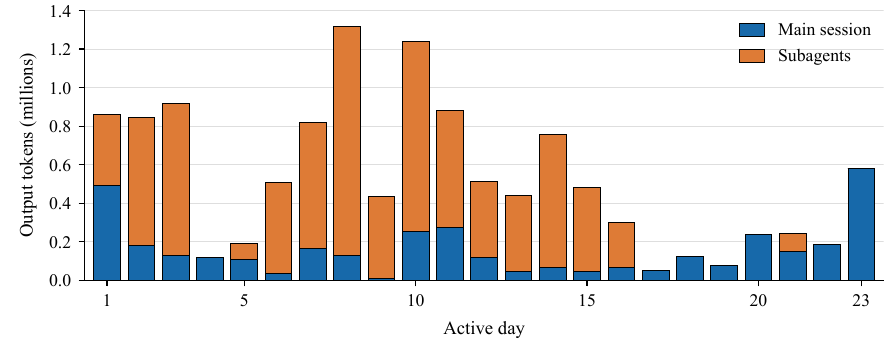}
  \caption{Output tokens per active workflow day.}
  \label{fig:workflow-output-tokens}
  \Description{Stacked bars show generated tokens for 23 active workflow days. Blue denotes the main session and orange denotes subagents. Output is concentrated in the earlier translation and review stages, with a peak of approximately 1.32 million tokens on day 8. Across the workflow, the main session generated 3.65 million tokens and subagents generated 8.48 million tokens.}
\end{figure}

Figure~\ref{fig:workflow-output-tokens} shows daily output token counts over 23 active days of agent-assisted NICAM implementation, overall code review and revision, and tasks within the three-tier verification protocol, such as developing unit tests and generating ensemble test kits.
Together, these stages produced 12.13 million output tokens.
Subagent output was concentrated in the earlier implementation and review stages.
The main session handled most of the later work on verification, test kits, and ensemble analysis.

We did not record human working hours, so we asked Claude Code to count the instructions sent to the agent during these active days.
Table~\ref{tab:human-effort} summarizes the instruction counts, median lengths, and assistant messages per instruction for each stage.
The 797 human instructions were accompanied by 6,007 assistant messages, averaging 7.5 messages per instruction.
Instruction lengths include pasted code and logs.
Days spent solely on human review or cluster work are outside this record.

\begin{table}[t]
  \caption{Human instructions and assistant messages across the workflow stages.}
  \label{tab:human-effort}
  \small
  \renewcommand{\arraystretch}{1.1}
  \setlength{\tabcolsep}{4pt}
  \begin{tabularx}{0.8\linewidth}{@{}>{\raggedright\arraybackslash}Xrrrr@{}}
    \toprule
    Stage & \shortstack[r]{Human\\instructions} &
      \shortstack[r]{Median length\\(words)} &
      \shortstack[r]{Assistant\\messages} &
      \shortstack[r]{Assistant messages\\per instruction} \\
    \midrule
    Porting progress & 355 & 34 & 2\,977 & 8.4 \\
    Overall code review/revision & 159 & 37 & 932 & 5.9 \\
    Verifications, 20-member ensemble & 169 & 20 & 1\,149 & 6.8 \\
    100-member ensemble & 114 & 15 & 949 & 8.3 \\
    \bottomrule
  \end{tabularx}
\end{table}

\subsection{Discussion}
Claude Code implemented all C++/Kokkos model code in this port, while the human directed the work and reviewed the results.
We believe the porting process went smoothly not only because of the capabilities of the LLM and coding agent but also because of NICAM's consistent data structures.
Most model data are stored in arrays organized as \texttt{(ij, k, g, l)}.
The Fortran source is well maintained and clearly structured, and its OpenACC and OpenMP directives provided useful guidance for the coding agent.
In our experience, the size of the codebase posed a greater challenge for human review than for the agent's translation work.

The three verification tiers provide complementary evidence of consistency with the Fortran reference for the tested configuration.
The surface-temperature initialization error illustrates why human review and integrated model runs remain necessary even after component-level tests pass.
Supplying a routine with reference inputs can bypass errors in the code that produces those inputs during a coupled simulation.
The current unit tests also use a limited set of dumped inputs and do not systematically cover boundary cases.
A broader collection of reference inputs is therefore needed to improve test coverage.

The performance results show that there is still room for architecture-specific tuning of the shared implementation.
Both Kokkos builds outperform the Fortran/OpenACC reference on GH200, although the MI300A run is slower than the GH200 run despite its higher peak memory bandwidth and FP64 throughput.
Using peak memory bandwidths of 5.3\,TB/s for MI300A and 4.02\,TB/s for GH200, the bandwidth-normalized kernel performance of the HIP build is approximately 96\% of that of the Fortran/OpenACC reference.
The HIP build currently retains the CUDA tile sizes, so tuning the execution policy for MI300A is one direction for further work.
Further profiling is needed to assess whether multidimensional index calculations, remaining conditional branches, or implementation choices made for readability contribute to the performance gap.

Because this was our first application-scale port with a coding agent, we frequently audited progress and requested confirmation before continuing.
In retrospect, we believe some of these exchanges could have been avoided.
The instruction counts therefore reflect both the work required for the port and our cautious approach to directing the agent.

Human review and the subscription's five-hour usage limits were the main constraints on progress.
The agent could generate code faster than the human could inspect it, so reviewing numerical logic and integration details took substantial time.
Token consumption was also difficult to predict for individual milestones, and development was repeatedly interrupted when the usage limit was reached.
Interruptions during implementation raised concerns about whether the agent could resume with sufficient context.
The completed port demonstrates the feasibility of agent-assisted modernization, with progress shaped by the human's capacity to review and verify the generated code.

\section{Conclusion}
\label{sec:conclusion}
We ported a production configuration of NICAM directly from Fortran/OpenACC to C++/Kokkos using Claude Code under human direction.
The port covers the dynamical core and seven physics packages, translating approximately 140,000 lines of Fortran into 35,000 lines of C++/Kokkos model source.
Validation combined component-level reference comparisons, short integration checks, and an ensemble-inspired consistency test.
Two 100-member ensembles provided evidence of consistency with the Fortran reference across 59 evaluated quantities over 10-day simulations.

Compared with the Fortran/OpenACC baseline on GH200, the port achieved application speedups of $1.71\times$ on GH200 and $1.54\times$ on MI300A.
The corresponding GPU kernel-time speedups were $1.38\times$ and $1.27\times$ through the CUDA and HIP backends, respectively.
After normalization by peak memory bandwidth, the HIP build reached 96\% of the Fortran baseline's kernel performance.
The overall recorded token usage corresponds to an estimated API list-price cost of USD~5,589.
These results demonstrate the feasibility of agent-assisted modernization at application scale with human direction and layered verification.

Future work will extend the port to additional NICAM components and broaden the reference inputs used in unit tests.
We also plan to use the development and verification procedures from this port to build reusable agent skills and a more automated workflow.
We are preparing a second porting exercise using the dry dynamical core of the current NICAM version to evaluate this workflow.

\begin{acks}
This work is partially supported by MEXT as “Project for Establishment of a Center for Advanced HPC-AI Development Support”. This research is partially supported by Exploratory Collaborative Research Project, Information Technology Center, The University of Tokyo. GPT-6 Astra was used solely for grammar checking and language refinement. All suggested revisions were reviewed by the authors, who take full
responsibility for the manuscript's content.
\end{acks}

\bibliographystyle{ACM-Reference-Format}
\bibliography{references}

\appendix
\section{The Human Instructions}
\label{sec:humaninst}
Listing~\ref{lst:kickoff} reproduces the supplied \texttt{kickoff.md} document in full.
Its Markdown notation and wording are retained, with long lines wrapped for readability.
\begin{lstlisting}[style=projectdocument,
  caption={Human kickoff instructions.},
  label={lst:kickoff}]
# Kickoff document for this project
- Final target: porting the whole Fortran NICAM to Kokkos

# Tasks
- Discuss the porting plan first, and write it to @Porting_Progress.md
- After confirming the plan with the user, develop a minimum executable Kokkos version of NICAM.

# Preferences & Requirements
- The development environment on the local desktop can be shown with ```module list```
- Activate the Python virtual environment with: ```source ~/.pyvenv/default/bin/activate```
- Build and execute NICAM through the docker environment
- Be careful with endianness: the Fortran NICAM uses big-endian I/O
- In the Kokkos version, NICAM's configuration file (nhm_driver.cnf) should use YAML instead

# RULES
- DO NOT put any files in /tmp
- Files outside the current project are READ-ONLY
- If you need extra libraries or software, ask the user to install them
- Run binaries one at a time: the local desktop has a single GPU, and concurrent runs conflict with each other. (This does NOT mean single-rank execution — MPI is kept in the port from the start.)
\end{lstlisting}

\section{The Porting Plan Generated by Claude Code}
\label{sec:portplan}
Listing~\ref{lst:portplan} presents essential excerpts from the initial \texttt{Porting\_Progress.md} generated on July~6, 2026.
Original wording is retained, with omissions marked by \texttt{[...]}.
These excerpts document the proposed strategy and milestones, which were subsequently refined during the project.
\begin{lstlisting}[style=projectdocument,
  caption={Selected excerpts from the initial agent-generated porting plan.},
  label={lst:portplan}]
# NICAM → Kokkos Porting: Plan & Progress
[...]
Final target (doc/kickoff.md): port the whole Fortran NICAM to C++/Kokkos.
Strategy: **standalone C++ rewrite** in `nicam-kks/`, milestone-by-milestone, each step validated against the Fortran reference before proceeding. First executable goal: **dry dynamical core** (physics off) reproducing the Fortran run on the real gl07rl00Az78 dataset.
[...]
### M1 — nicam-kks scaffold
[...]
**Exit**: smoke binary runs on Serial/OpenMP/CUDA; parses YAML; reads vgrid78.dat matching Fortran log; reads an init .nc matching `cdo info`.

### M2–M10 — dry core, dependency order
[...]
- **M2** ADM (rl00-prc01.info topology, index helpers) + CNST + calendar/time → table/constant diffs
- **M3** GRD: vgrid read, hgrid/topo NetCDF, GRD_scaling → Stage-A diffs
- **M4** COMM (before operators — GMTR/OPRT setup needs it): copy-list construction + device-side transfer → table diffs + synthetic index-encoded field test
- **M5** GMTR + OPRT/OPRT3D coefficient setup + VMTR (incl. `_pl` and singular points) → Stage-A diffs; GTL area sums vs log
- **M6** OPRT kernels (divergence/gradient/laplacian/diffusion/horizontalize) → Stage-D in/out diffs, CPU then CUDA
- **M7** prgvar init read + thrmdyn/cnvvar/bndcnd/bsstate + **restart writer, step-0 round-trip vs init files** (de-risks NetCDF writer early) → Stage-C diffs
- **M8** mod_src + mod_numfilter (NONLINEAR1 hdiff has min/max reductions: compare at ~1e-12) → Stage-D diffs
- **M9** vi_small_step (tridiagonal solver — keep identical k sweep order) + tracer advection (MIURA2004 + Thuburn limiter branch semantics) → Stage-D diffs
- **M10** full dynamics_step + driver + minimal history. [...]
- **M11** CUDA hardening; nsys baseline vs OpenACC kernel timings (oprt_diffusion, oprt_divdamp, vi_*)

### P0–P5 — physics (sketch; each package gets its own Fortran reference run with only it enabled)
- **P0** physics infrastructure: sfcvar/extdata (time interpolation, periodic year)/radvar, physics_step skeleton all-NONE → dry results unchanged
- **P1** NSW6 microphysics + precip transport (KIN2EIN) — column-local, no COMM
- **P2** MYNN2 turbulence + Louis surface (prescribed SST)
- **P3** MSTRNX radiation (largest single module; PARA.bnd29ch111sp big-endian tables, ozone extdata)
- **P4** MATSIRO land + mixed-layer ocean + MCFARLANE GWD + sfc_step coupling
- **P5** full physics vs `run/reference/full-gpu`; then longer runs with budget monitors
[...]
## 5. Decisions log
[...]
- **2026-07-06** YAML config via vendored fkYAML single header (user-confirmed).
[...]
- **2026-07-06** Double precision (matches Fortran build: no -DSINGLE).
[...]
- **2026-07-06** Fortran tree is add-only (dump harness as new files); no edits to existing modules.
[...]
\end{lstlisting}

\end{document}